\documentstyle[epsfig]{aipproc}
\begin{document}
\title{Modern Supernova Search} 

\author{Myung Gyoon Lee} % $^*$}
\address{Astronomy Program, SEES, Seoul National University.\\
Seoul 151-742, Korea \\
Email: mglee@astrog.snu.ac.kr}

\lefthead{MODERN SUPERNOVA SEARCH}
\righthead{Lee}
\maketitle

\begin{abstract}
Supernovae play a critical role in observational cosmology as well as 
in astrophysics of stars and galaxies. 
Recent era has seen dramatic progress in the research of supernovae.
Several programs to search systematically supernovae in nearby to distant 
galaxies have been very successful.
Recent progresses in the modern supernova search are reviewed.
\end{abstract}

\section*{Introduction}

Ancient supernova (SN) searches(?) must have been based on random visual observations and
discovered eight SNe all of which are in our Galaxy.
It is almost 400 years since the last galactic supernova SN 1604 was observed. 
Korea has a rich record of old observations of transient objects including 
galactic supernovae which were called as `guest stars' before (e.g. see Chu 1968).
Fig. 1 illustrates a light curve of SN 1604 based on the old Korean and European observations
(Clark \& Stephensen 1977). A light curve of a Type Ia Supernova 1991T 
(Lira et al. 1998) is also
overlayed in Fig. 1, arbitrarily shifted to match SN 1604 around the peak. 
The light curve of SN 1604 is approximately matched by that of the Type Ia, showing
 how remarkable the old visual photometry of SN 1604 was even at that time.

\begin{figure}[t!] % fig 1
\centerline{\epsfig{file=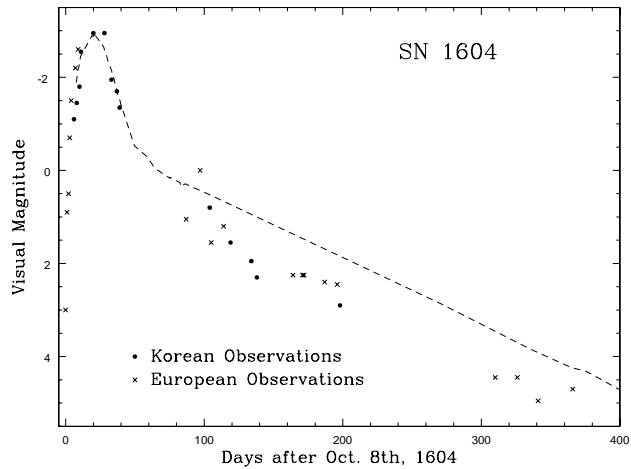,height=3.5in,angle=270}} %width=3.5in}}
\vspace{10pt}
\caption{A light curve of SN 1604 based on the Korean (filled circles) and European (crosses) records
(Clark \& Stephenson 1977).
The dashed line represents a $B$-band light curve of Type Ia SN 1991T (Lira et al. 1998).
 Note how remarkable the old visual photometry of SN 1604 was even at that time.}
\label{fig1}
\end{figure}

In 1885 the first extragalactic supernova (SN 1885) was discovered in M31 after 281 years passed
since the last galactic supernova was discovered.
However, it is only in 1934 that the concept of supernovae was first introduced by Baade \& Zwicky (1934)
and the first systematic supernova (SN) search was made by Zwicky (1938). 
Since then many supernova searches were conducted.

There are several reasons for doing SN search. First, it is a lot of fun to discover
new objects in the sky (which is a very important motivation for amateur astronomers).
Secondly, SN searches provide diverse informations of the astrophysics of stars and galaxies.
Thirdly, SN searches can be also used for observational cosmology, especially to determine
the cosmological parameters (see a recent review by Leibundghut (2000)).

However it is only in 1990's that systematic SN searches using modern instruments were conducted,
and it is about time for SN search to bloom today.
As many as about 1750 extragalactic SNe were discovered as of July, 2000 (up to SN 2000cz).

\section*{Recent Supernova Searches}

Modern SN searches are characterized by either fast automatic search with small telescopes 
covering a wide field for nearby SNe and other transient objects such as gamma ray bursts (GRBs),
or deep search in a small field with large (4m class) telescopes 
well scheduled for maximizing the SN discovery rate.

Recent professional SN searches can be divided into three kinds: the high-z SN searches, 
the low-z SN searches, and the cluster SN searches.
Among the high-z SN searches are the High-z SN search team (HzSS; Garnavich etal. 1998) and 
the SN Cosmology Project (SCP; Perlmutter et al. 1999). 
Among the low-z SN searches are 
the Lick Observatory SN Search (LOSS; Filippenko et al. 2000),
the Beijing Astronomical Observatory SN Survey (BAOSS; Li et al. 1996, 2000),
the EROS Nearby SN Search (EROSNSS; Hardin etal. 2000),
the Nearby Galaxies SN Search (NGSS; Strolger et al. 2000),
the QUEST (Schafer et al. 1999, Schafer 2000),
%the NEAT
the European SN Cosmology Consortium (ESCC; Hardin et al. 1999), and
the Super Livermore Optical Transient Imaging System (Super-LOTIS; Williams et al. 2000, Park 2000).
Among the cluster SN searches are
the Mount Stromlo Abell Cluster SN Search (MSACSS; Reiss etal. 1998),
the Wise Observatory Optical Transients Search (WOOTS; Gal-Yam \& Maoz 1998), and
the Seoul National University SN Search (SNUSS; Lee et al. 1999a,b).
Amateur astronomers also have played important roles in SN discovery 
(e.g. Evans (2000), the UK nova/SN patrol (Armstrong 2000), 
the Tenagra Observatories (Schwartz 2000), and the Pucket Observatory (Pucket 2000)). 

Table 1 lists the number of SNe discovered by some of these SN search teams 
for the last three years (1998 - July 29, 2000). 
 It shows that the recent SN searches have been very successful. 
Outstanding examples are SCP and HzSS for high-z SNe, and LOSS, EROSNSS and QUEST for low-z SNe.

\begin{table}
\caption{A List of Recent SN Searches }
\label{table1}
\begin{tabular}{cccccl} %\begin{tabular}{lddd}
Team & N(1998)$^a$ & N(1999)$^a$ & N(2000)$^a$ & 
N(total)\tablenote{Numbers of SNe discovered for 1998--July 29, 2000.}  & Remarks \\
\tableline
\multicolumn{6}{c}{High-z SS} \\ %\tablenote{Estimated 2.}(\%)}\\
\tableline
SCP & 23 & 43 & 0 & 66 & 4m \\
HzSS & 20 & 33 & 0 & 53 & 4m \\
\tableline
\multicolumn{6}{c}{Low-z SS} \\
\tableline
LOSS  & 19 & 40 & 15 & 74 & 0.7m, automatic \\
EROSNSS & 13 & 22 & 14 & 49 & 1m \\
QUEST & 0 & 0 & 33 & 33 & 1m Schmidt, scanning, $2.3^\circ \times 15^\circ $ per hour \\
NGSS & 0 & 14 & 8 & 22 & 1m, $59'\times 59'$ \\
BAOSS & 8 & 5 & 0 & 13 & 0.6m, China \\
%NEAT & 0 & 5 & 0 & 5 & \\
ESCC & 0 & 12 & 0 & 12 & 1m-4m \\
\tableline
\multicolumn{6}{c}{Cluster SS} \\
\tableline
MSACSS & 22 & 2 & 0 & 24 & 1m, Australia \\
WOOTS & 5 & 12 & 0 & 17 & 1m, Israel \\
SNUSS & 0 & 1 & 0 & 1 & 1.8m, test, Korea \\
\end{tabular}
\end{table}

\begin{figure}[h!] % fig 2
\centerline{\epsfig{file=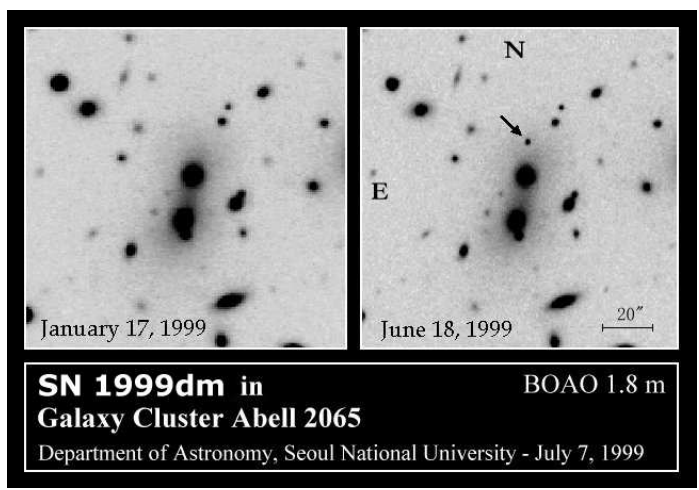,height=2.5in}} %,width=3.5in}}
\vspace{10pt}
\caption{Finding charts for SN1999dm in Abell 2065 discovered by the SNUSS.}
\label{fig2}
\end{figure}

Fig. 2 shows an example of SNe in the Abell clusters discovered by the SNUSS (Lee et al. 1999a, b), 
SN 1999dm in Abell 2065. 

Nearby clusters of galaxies are ideal places for SN search,
because there are included many galaxies in a single field covered by a telescope. It is expected
that many more SNe in the clusters of galaxies will be discovered soon.
Saini et al. (2000) suggested an interesting idea of searching for SNe in the gravitationally lensed
arclets in distant clusters, because those SNe are expected to be magnified by up to 3-4 magnitudes.

\section*{Supernova Statistics}

A recent SN catalog was given by Barbon et al. (1999), and updated SN catalogs 
are provided in CBAT (http://cfa-www.harvard.edu/iau/lists /Supernovae.html) and http://merlino.pd.astro.it/$~$supern/. Various informations of SNe are also 
available in the International Supernova Network (http://www.supernovae.net/isn.htm).
The recent SN catalogs show that there are 1747 SNe discovered until July 29, 2000 (up to SN 2000cz).

\begin{figure}[h!] % fig 3
\centerline{\epsfig{file=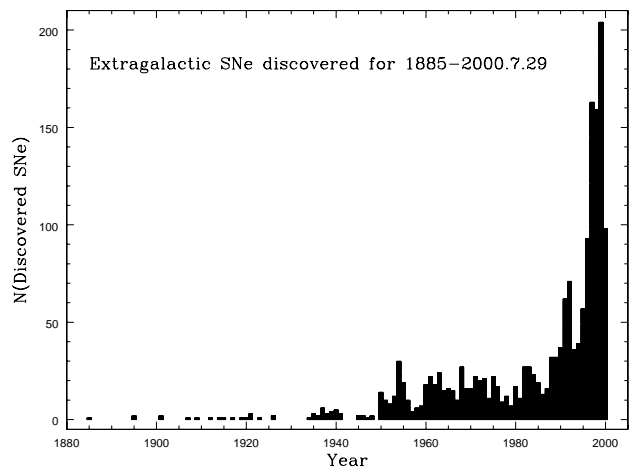,height=3.5in, angle=270}} %width=3.5in}}
\vspace{10pt}
\caption{Number distribution of the supernovae discovered for 1885 -- July 29, 2000.}
\label{fig3}
\end{figure}

Fig. 3 displays the  number distribution of SNe discovered 
for 1885 -- July 29, 2000.
It shows that the number of discovered SNe started increasing steeply in 1990's. 
100 SNe per year was broken in 1997 and 200 SNe per year was broken in 1999. 
It is expected that this
trend of increasing may continue for a while in the beginning of 21th century.

The SN types are known for about 1080 SNe: 540 Ia+Iap (50\%), 291 II+IIp (27\%), 95 I+Ip (10\%),
71 Ib+Ic+Ib/c (6\%), 48 IIn+IIb (4\%), 36 IIn+IIb (3\%). 
Most of host galaxies are spiral galaxies. Only I+Ip and Ia+Iap are discovered in elliptical galaxies,
while no I+Ip and Ia+Iap are discovered in irregular galaxies.

There are several galaxies with multiple SNe. Two of them (NGC 5236 and NGC 6946) have six SNe each,
and three (NGC 2276, NGC 3690, NGC 4321) have four SNe each.
The faintest SN  is SN 1999FF with $I\sim 27$ mag in the Hubble Deep Field 
(Gilliland et al. 1999) and the most distant
SNe is at z=1.2 (SN 1999fv) or 1.3 (SN 1997ff).

\section*{Highlights of Recent Supernova Searches}

There are many important new results emerged from recent SN searches, a few of which
are listed below (but see also the remaining problems pointed out by Suntzeff (2000)). 

(1) The most exciting news among the recent results in the SN searches may be 
a discovery by two high-z SN search teams that the universe may be accelerating,
with $\Omega_{\Lambda}\sim 0.7$ and $\Omega_M \sim 0.3$
(Perlmutter et al. 1999; Garnavich et al. 1998). 
Fig. 4 (Riess 2000) illustrates  the Hubble diagram of distant SNe Ia discovered by the SCP and the HzSS 
as well as nearby SNe Ia discovered by the C\'{a}lan/Tololo Survey (Hamuy et al. 1996).
Although the present data provide reasonable evidence for the acceleration universe, 
better data for the high-z SNe are still needed to derive a definitive conclusion on the
linear systematic effect due to evolution or grey extinction. 

\begin{figure}[h!] % fig 4
\centerline{\epsfig{file=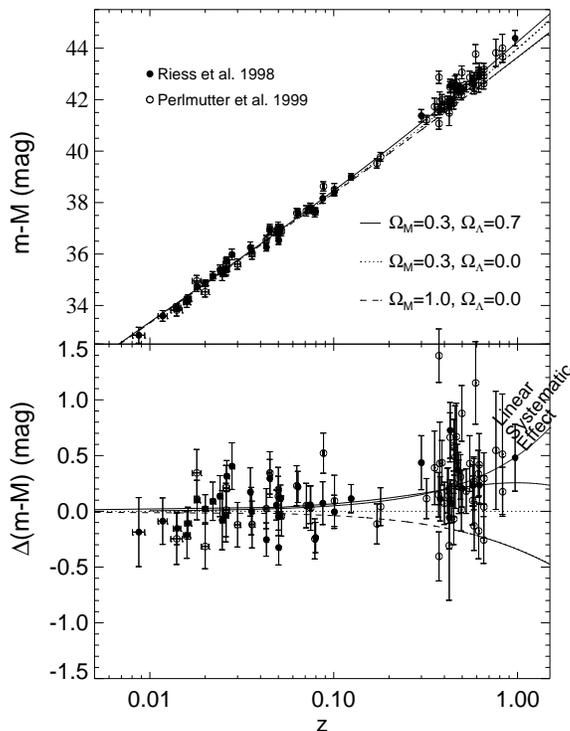,height=3.5in}} %,width=3.5in}}
\vspace{20pt}
\caption{Upper panel: the Hubble diagram of SNe Ia at low to high-z.
%Perlmutter et al. (1999; SCP), Riess et al. (1998; HzSS),and Hamuy et al. (1996; C\'{a}lan/Tololo Survey).
Lower panel: the difference between data and models from the $\Omega_M = 0.3$ and
$\Omega_{\Lambda} = 0.7$ prediction, showing also the effect of a systematic error which grows
linearly with redshift (Riess 2000).}
\label{fig4}
\end{figure}

(2) Two SNe (SN 1997ff, SN 1997fg) were discovered in the Hubble Deep Field North 
from the images taken with an interval of two years (Gilliland et al. 1999). 
The redshift of SN 1997ff determined photometrically is z = 1.32 ($I=27.0$ mag). This is the most
distant SN among the known SNe, if confirmed spectroscopically.   

(3) The Hubble constant was determined with improved accuracy using the SN Ia in the galaxies
to which Cepheid distances were measured with the Hubble Space Telescope (HST).
The HST $H_0$ team presented a value of $H_0 = 68\pm7$ km s$^{-1}$ Mpc$^{-1}$ (Gibson et al. 2000),
 and the HST SN team presented a value of $H_0 = 59\pm6$ km s$^{-1}$ Mpc$^{-1}$ (Parodi e al. 2000).
There is still some difference between the two estimates, but they are much closer than before.

(4) Recently it has been suggested that SN Ib/c may be linked with the Gamma-Ray Burst (GRB):
SN 1998bw/GRB980425 and SN 1997cy/GRB970514, but it
needs further studies to confirm it (Galama et al. 1998, Iwamoto et al. 1998, 
Paczy\'{n}ski 1999, Brown et al. 2000). 

\section*{Future}

It is only a decade that modern supernova searches started. It is very impressive and encouraging
that recent supernova searches made a great progress in such a short period of time.
However, even more impressive and revolutionary results may come  from the supernova searches 
in the 21th century.
There are two directions in the future for the supernova search: all-sky search and deep-sky search.
All-sky search is to search for nearby SNe in the entire sky using fast wide field equipments, and
deep-sky search is to search for distant SNe in a small field using  ground-based large telescopes
or space telescopes (e.g. see a supernova pencil beam survey suggested by Wang (2000)).

One space mission to be exclusively used for supernova search is being planned by Perlmutter et al.
(2000): SN Acceleration Probe Satellite (SNAP). SNAP will use a wide-field 2m class telescope, 
covering one degree$^2$ field of view, with a primary aim of pinning down the cosmological
parameters with Type Ia SNe. 
It is expected that about 2000 SNe Ia at $z<1.7$ will be discovered every year
for three years using the SNAP. With these data, the linear effect as shown in the lower panel of 
Fig. 4 can be tested.
More distant supernovae at $z<5-8$ in a small field will be searched in the space 
using the Next Generation Space Telescope (NGST) with an 8m aperture, which is scheduled
to be launched in 2009.
Even more distant supernovae at $z<10$, if any, will be discovered using the 100m-class ground-based 
optical telescopes. For example, a conceptual study of the OverWhelmingly Large Telescope (OWL)
with the 100m aperture is being done now (Gilmozzi \& Dierickx 2000). 
With the results to come, the theoretical predictions of the cosmic SN rate history 
(Kobayashi et al. 2000,  Dahl\'{e}n, \& Fransson 2000) can be tested.
A new golden age era is coming soon in the supernova search.

%\landscape

\end{document}